\documentclass[floatfix, plb, nopacs, twocolumn,showkeys, preprintnumbers, nofootinbib, superscriptaddress]{revtex4-1}
\usepackage[utf8]{inputenc}
\usepackage[sort&compress]{natbib}
\usepackage{ulem}
\usepackage{overpic}
\usepackage{bm}
\usepackage{times}
\usepackage{amssymb,amsbsy,amsmath,amsfonts}
\usepackage{graphicx}
\usepackage{float}
\usepackage{color}
\usepackage{booktabs}
\usepackage{makecell}

\usepackage{rotating}
\usepackage{srcltx}
\usepackage{slashed}
\usepackage{subfigure}
\usepackage{multirow}
\usepackage{verbatim}
\usepackage{hyperref}
\usepackage{tabularx}

\DeclareUnicodeCharacter{3002}{HEREHEREHERE}

\begin{document}

\title{Predicting reaction observables for the two-neutron halo candidates $^{31}$F and $^{39}$Na}

\author{Jia-Lin An}
\affiliation{School of Physics, Beihang University, Beijing 100191, China}
\affiliation{Baoding Hospital of Beijing Children's Hospital, Capital Medical University, Baoding 071000, China}

\author{Li-Yang Wang} 
\affiliation{School of Physics, Beihang University, Beijing 100191, China}

\author{Kaiyuan Zhang}\email{zhangky@caep.cn}
\affiliation{National Key Laboratory of Neutron Science and Technology, Institute of Nuclear Physics and Chemistry, China Academy of Engineering Physics, Mianyang, Sichuan 621900, China}

\author{Shi-Sheng Zhang}\email{zss76@buaa.edu.cn}
\affiliation{School of Physics, Beihang University, Beijing 100191, China}

\date{\today}

\begin{abstract}
Microscopic description of two-neutron ($2n$) halo candidates $^{31}$F and $^{39}$Na has been realized from nuclear structure to reaction observables for the first time.
The reliability of the Glauber reaction model has been confirmed by exactly reproducing the momentum distributions of the benchmark $2n$ halo nucleus $^{11}$Li, with the identical structural inputs from the former work.
Combined with the structure from the deformed relativistic Hartree-Bogoliubov theory in continuum (DRHBc), the Glauber model is applied to predict the reaction observables, including the reaction cross sections (RCSs) for the fluorine and sodium isotopes bombarding a carbon target at 240~MeV/A and the longitudinal momentum distributions of the fragments after $2n$ knockout reactions.
It turns out that the calculated RCSs agree well with the available experimental data and a pronounced increase occurs to $^{29, 31}$F + $^{12}$C and $^{37, 39}$Na + $^{12}$C, which deviate from the original trend of their neighbours.
Furthermore, the narrower longitudinal momentum distributions of the fragments after $2n$ knockout reactions demonstrate that $^{31}$F and $^{39}$Na have the dilute $2n$ halo structure.
Such a new combination is promising to suggest new $2n$ halo candidates for future measurements.
\end{abstract}

\keywords{$2n$ halo, $^{31}$F, $^{39}$Na, the Glauber Model, the DRHBc theory, reaction cross section, longitudinal momentum distribution}

\maketitle

\section{Introduction}\label{sec:Introduction}

Radioactive ion beams enable the study of exotic nuclei far from the stability valley. 
Those nuclei often exhibit novel phenomena, such as halo~\cite{al2004halo} and nucleon emission, etc. 
Since the formation of halo nuclei is mainly caused by the deformation and the population of the bound and resonant states around the Fermi surface in the loosely bound system, new magic numbers or $``$ island of inversion $"$~\cite{Nakamura2009PRL, Nakamura2014PRL} might occur, which played an important role in the identification of halo nuclei $^{31}$Ne and $^{37}$Mg in the last decade.

Halo nucleus is commonly characterized by a dilute spatial distribution of valence nucleon(s) in a weakly bound system with small separation energy. 
Experimentally, reaction observables are regarded as direct evidences for a halo nucleus, such as a pronounced increase of the reaction cross sections (RCSs) and a narrow longitudinal momentum distribution of the residues after valence neutron(s) knockout reactions~\cite{Jensen2004RMP, hansen1987EPL, Ring1996PPNP, Zhang2014PLB, FangDQ2016NST}. 

The first discovered halo nucleus $^{11}$Li consists of two dilute valence neutrons and a tightly bound core nucleus $^{9}$Li, which exhibited a topological Borromean structure and opened a new research field~\cite{Tanihata1985PRL, hansen1987EPL, Kobayashi1988PRL}. 
In a recent experiment, the unexpectedly large RCSs and derived matter radius identified $^{29}$F as the heaviest two-neutron ($2n$) Borromean halo nucleus~\cite{Bagchi2020PRL}. 
Sooner or Later, new isotopes $^{31}$F~\cite{Ahn2019PRL} and $^{39}$Na~\cite{Ahn2022PRL} have been discovered at the RIKEN Radioactive Isotope Beam Factory by using the projectile fragmentation of an intense $^{48}$Ca beam on a beryllium target at 345~MeV/A.
These measurements cast a new light on the search of heavier $2n$ halo candidates and contribute to a better understanding of novel nuclear structure under extreme neutron-rich conditions.
Except for the binding energies (or two-neutron separation energy $S_{2n}$) and the RCSs, no data are available for the momentum distributions of the fragments after the breakup reactions for the neutron-rich fluorine and sodium nuclei, which are the most direct and decisive evidence to identify as a $2n$ halo nucleus or not. Therefore, a theoretical study from the microscopic structure to the reaction observables is desirable and will support the future measurements.

So far, some theoretical studies have been done on $^{31}$F and $^{39}$Na~\cite{singh2022PRC, Chai2020PRC}.
Singh \emph{et al.} systematically investigated the ground-state configuration mixing, matter radius, and dipole response of $^{31}$F by employing a three-body formalism in hyperspherical coordinates and the transformed harmonic oscillator (THO) basis method, in which the core-neutron interactions are adjusted to reproduce different structural scenarios. 
Their results indicate that $^{31}$F exhibits a significant spatial extension and enhanced dipole strength under configurations dominated by the $2p_{3/2}$ component, which strongly support its $2n$ halo structure~\cite{singh2022PRC}.
Meanwhile, much attention has been focused on $^{39}$Na because of the quenched $N=28$ shell closure and shape decoupling. 
Within the deformed relativistic Hartree-Bogoliubov theory in continuum (DRHBc)~\cite{Zhou2010PRC(R),Li2012PRC,Zhang2020PRC,Pan2022PRC}, Zhang {\itshape et al.} found that the lowering of $2p$ orbitals in the spherical limit results in the collapse of the $N=28$ shell closure for $^{39}$Na, which has a well-deformed ground state~\cite{Zhang2023PRC(L1)}. 
These works are still confined in the theoretical study on the structure, lack of the bridge between the predictions and the reaction observables.

Lately, we combined the microscopic structure theory with the Glauber reaction model to describe one-neutron ($1n$) halo nuclei, such as $^{15,19}$C~\cite{wang2024EPJA,An2025JPG}, the heavier $^{31}$Ne~\cite{Zhang2022JPG,Zhong2022SciChina} and the heaviest $^{37}$Mg~\cite{Zhang2023PLB,An2024PLB}. 
Since it is even harder to search for heavier $1n$ halo nucleus from the experiment compared to $2n$ halo case~\cite{marques2000PLB}, we aim at extending the DRHBc + Glauber approach to more complex $2n$ halo system in this Letter. 

As the first step, the reliability of the Glauber model for the $2n$ halo case has been demonstrated by accurately reproducing the well-established reaction observables for the $2n$ halo nucleus $^{11}$Li bombarding a carbon target.
Then, we replace the structure inputs for the Glauber model with those from the DRHBc theory, and apply this scheme to systematically study the reaction observables, including the RCSs and the longitudinal momentum distributions of the residues after $2n$ removal reactions for the neutron-rich fluorine and sodium isotopes on a carbon target.

This Letter is organized as follows.
The theoretical formalism is briefly outlined in Section \ref{sec:Frame}.
Our results and detailed discussions are presented in Section \ref{sec:Results}. 
Finally, we make a summary in Section \ref{sec:Summary}.

\section{Theoretical Framework} \label{sec:Frame}

To build up the bridge from $2n$ halo structure to reaction observables, we combine the DRHBc~\cite{Li2012PRC,Zhang2020PRC} structure theory with the Glauber reaction model~\cite{abu2003, glauber, Horiuchi2010PRC, MaZY2001, Fang2004PRC}. 
In concise, the DRHBc theory solves the relativistic Hartree–Bogoliubov (RHB) equation in a Dirac Woods–Saxon basis~\cite{Zhou2003PRC,Zhang2022PRC}.
This basis yields a superior asymptotic wave function compared to the harmonic oscillator basis, which is crucial and suitable for the description of weakly bound system. Details can be found in Refs.~\cite{Zhou2003PRC,Zhang2022PRC} and references therein. The density functional PC-PK1~\cite{Zhao2010PRC} is used in the following DRHBc calculations.

Taking nuclear structure information from the DRHBc theory as the inputs of the Glauber reaction model, we simulate the scattering of the $2n$ halo nucleus on a carbon target, assuming the microscopic optical-limit approximation. 
Different from the $1n$ case, the RCSs $\sigma_{R}$ for a $2n$ halo projectile (P) on a target (T) are written as
\begin{align}
\label{reaction cross sections}
&\sigma_{R} 
=\int {\rm d}\bm{b}\times\\
\nonumber&(1-| \langle{\varphi_0}|e^{i\chi_{CT}(\bm{b}_C)+i\chi_{NT}(\bm{b}_C+\bm{s_1})+i\chi_{NT}(\bm{b}_C+\bm{s_2})}|{\varphi_0}\rangle|^2).
\end{align}
Here $\bm{b}$ denotes the impact parameter between the projectile and the target, and $\bm{s_i}$ refers to is the transverse coordinate of the $i$-th valence neutron relative to the core's center of mass. The wave functions of valence neutron $\varphi_0$ and the phase-shift function $\chi$ are decisive inputs from the structure. 

Meanwhile, the longitudinal momentum distribution of the residues - the core nucleus and two valence neutrons - from the breakup reaction can be calculated by the new expression
\begin{align}
\label{momentum distribution}
\nonumber&\frac{{\rm d}\sigma_{-N}^{inel} }{{\rm d}\bm{p}_{\parallel} } 
=\frac{1}{2\pi\hbar}\int {\rm d}\bm{s}e^{-2Im\chi_{CT}(\bm{b}_C)}\times\\
&\int {\rm d}\bm{b}_N\left[1-e^{-2Im\chi_{NT}(\bm{b}_C+\bm{s_1})}-e^{-2Im\chi_{NT}(\bm{b}_C+\bm{s_2})}\right]\times\\
\nonumber&\int {\rm d}z\int {\rm d}{z'}e^{\frac{i}{\hbar}\bm{p}_{\parallel}(z-{z'})}u_{nlj}^{1*}(r'_1)u_{nlj}^{1}(r_1)u_{nlj}^{2*}(r'_2)u_{nlj}^{2}(r_2)\times\\
\nonumber&\frac{1}{16\pi^2}P_l({\hat{\bm{r_1}}}\cdot{\hat{\bm{r'_1}}})P_l({\hat{\bm{r_2}}}\cdot{\hat{\bm{r'_2}}}),
\end{align}
where $P_l$ denotes the $l$-order Legendre Polynomial as a function of $\bm{r_{1,2}}=(\bm{s_{1,2}},z_{1,2})$ and $\bm{r'_{1,2}}=(\bm{s'_{1,2}},z'_{1,2})$.

It can be seen that these observables are mainly decided by the phase-shift functions, which reflect the information of the interactions.
The phase-shift functions ~$\chi_{CT}$ ($\chi_{NT}$) denote the interaction between core (whole) nucleus and the target as follows,
\begin{gather}
    i\chi_{CT} = \int q\rho_C(q) \rho_T(q) f_{NN}(q) J_0(qb) {\rm d} q, \\
    i\chi_{NT} = \int q\rho_T(q) f_{NN}(q) J_0(qb) {\rm d} q.
\end{gather}

\section{Results and discussion}\label{sec:Results} 

\subsection{Validity for $^{11}$Li bombarding a carbon target}

To check the validity of the Glauber model for $2n$ halo case, we firstly applied it to $^{11}$Li + $^{12}$C case~\cite{ogawa2001}. 
Same as Ref.~\cite{ogawa2001}, the density distributions for core nucleus are fitted by two-set of Gaussian functions (GFs), and the wave functions of $2n$ in a harmonic oscillator potential assuming two valence neutrons occupy the orbitals $(s_{1/2})^{2}$ and $(p_{1/2})^{2}$.
As listed in Table~\ref{tab:reac}, the RCSs agree well with the results in Ref.~\cite{ogawa2001} within the relative error of $1\%$.

\setcounter{table}{0}
\begin{table}[ht]
    \centering
    \caption{The RCSs for $^{11}$Li + $^{12}$C reaction compared with the results from Ref.~\cite{ogawa2001}.}
\begin{tabular}{m{13em} m{6em} m{7em}}
        \hline
        Valence Neutron Configuration & Theo. (mb) & This work (mb) \\
        \hline
        $(s_{1/2})^{2}$ & 1050 & 1051 \\
        $(p_{1/2})^{2}$ & 1060 & 1059 \\
        \hline
    \end{tabular}
    \label{tab:reac}
\end{table}

\begin{figure}[htbp]
    \centering
    \includegraphics[width=0.8\linewidth]{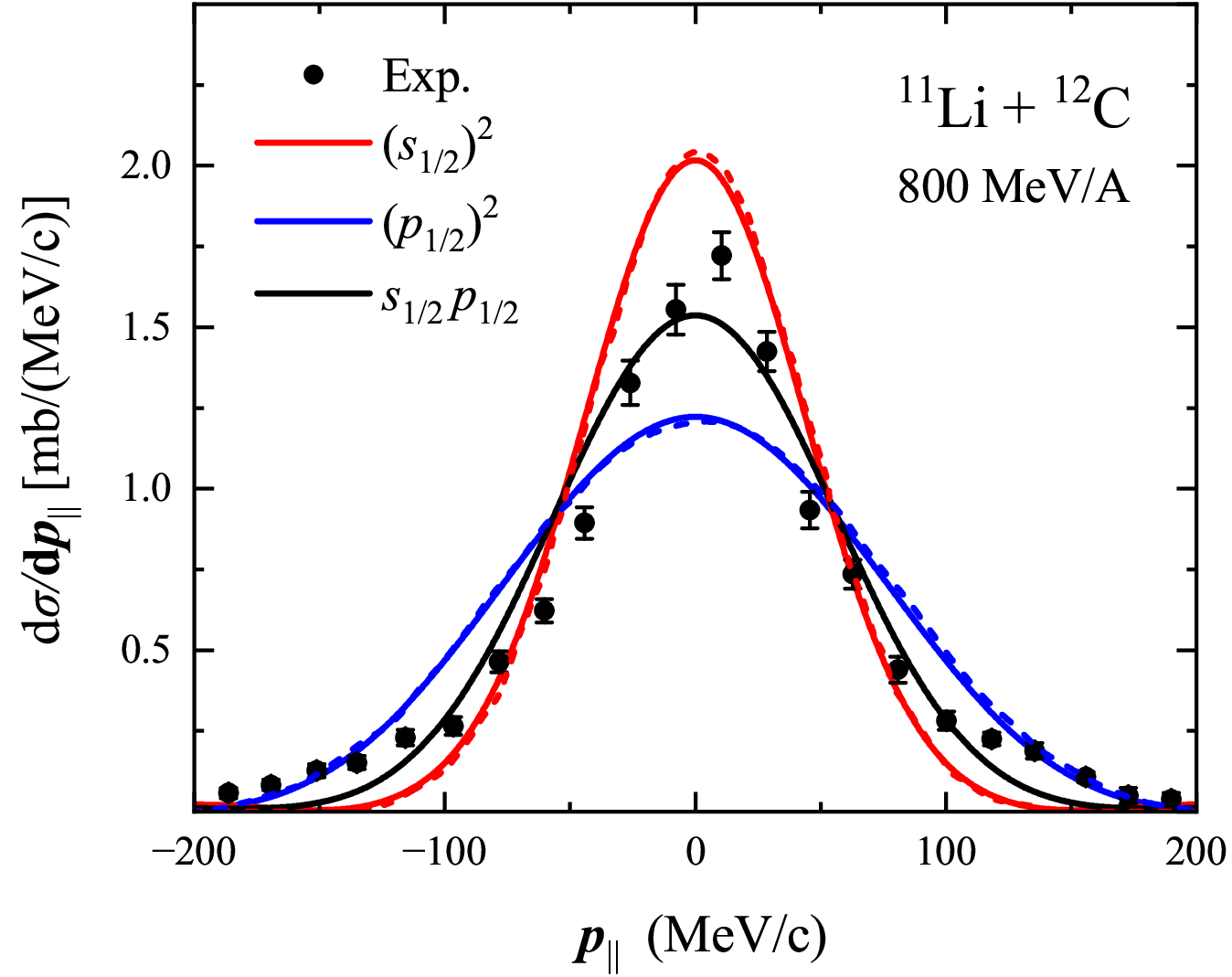}
    \caption{(Color Online) Inclusive longitudinal momentum distributions of $^{9}$Li residues after $2n$ removal reaction of $^{11}$Li + $^{12}$C at the incident energy 800~MeV/A. 
    The red and blue solid/dashed lines denote our/their~\cite{ogawa2001} predictions with the Glauber model for the configurations $(s_{1/2})^{2}$ and $(p_{1/2})^{2}$, respectively.
    The black solid line shows the results of mixed configuration $s_{1/2}p_{1/2}$.
    Experimental data (closed black circles) are taken from Ref.~\cite{Kobayashi1988PRL} for comparison.
}
    \label{fig:momdist_Li}
\end{figure}

Furthermore, we compute the longitudinal momentum distributions for the residues after $2n$ removal from reaction $^{11}$Li + $^{12}$C  at 800~MeV/A. 
As shown in Fig.~\ref{fig:momdist_Li}, our results (normalized to the experimental total cross-section) perfectly reproduce the profiles from the former work~\cite{ogawa2001}. 
The momentum distributions for the case of two valence neutrons occupying orbital $(s_{1/2})^{2}$/$(p_{1/2})^{2}$ exhibit typically narrow/broad shapes, which accurately reflect the spatial diffuseness of the wave functions for $2n$ on different orbital. 
In addition, our results for the mixed configuration $s_{1/2}p_{1/2}$ lies in between, consistent with the experimental data~\cite{Kobayashi1988PRL}.

The well-reproduced results confirm the effectiveness of the Glauber model for the reaction case of $2n$ halo nucleus.
Therefore, we can safely implant it into the reactions caused by the projectiles $^{31}$F and $^{39}$Na.

\subsection{Application to $^{31}$F on a carbon target}

\begin{figure}[htbp]
\centering
\includegraphics[width=0.58\linewidth]{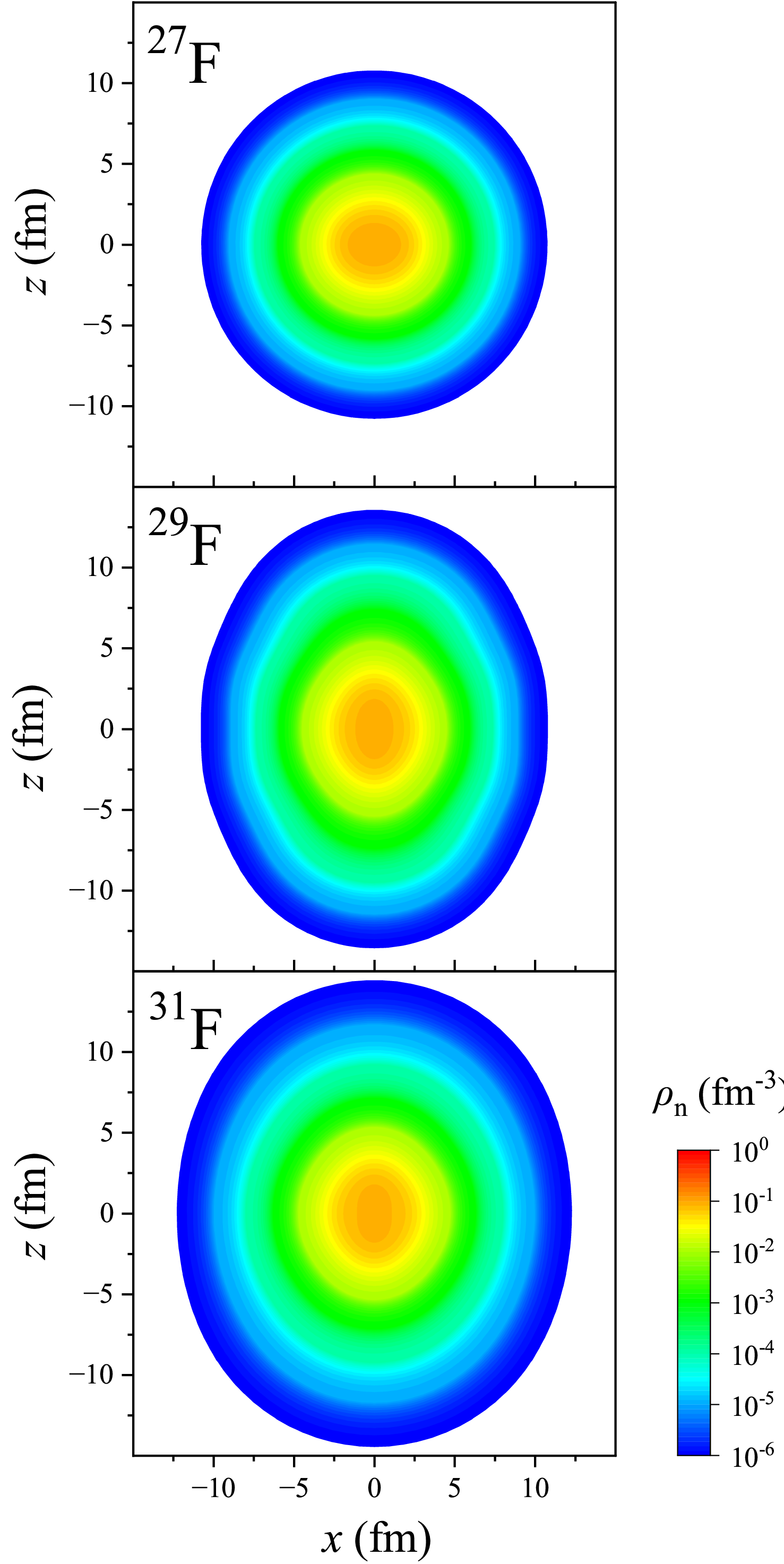}
  \caption{(Color online) Two-dimensional neutron density distributions for $^{27,29,31}$F in the $xz$ plane with the $z$ axis as the symmetry axis. 
  Here the results for $^{29}$F are obtained from a constrained calculation with a deformation parameter $\beta_2=$ 0.3.
  }
\label{fig:density_F}
\end{figure}

Based on the valid Glauber model for $2n$ halo nucleus $^{11}$Li, we replace the structure information by those for neutron-rich fluorine isotopes from the DRHBc theory. 
For $^{31}$F, the DRHBc theory predicts an extremely small $S_{2n}$ ($\approx 0.41$~MeV) , which is the first signal to have a halo structure.
Then, we plot two-dimensional neutron density distributions for $^{27,29,31}$F in $xz$ plane with the symmetry $z$ axis in Fig.~\ref{fig:density_F}.
Significant spatial extension of the density distributions in $^{29,31}$F can be clearly seen compared to those in $^{27}$F. 
Mentioned that the DRHBc calculations for $^{29}$F is constrained by a commonly proposed deformation parameter $\beta_2=$ 0.3~\cite{hamamoto2021PLB}.

\begin{figure}[htbp]
  \centering
  \includegraphics[width=0.8\linewidth]{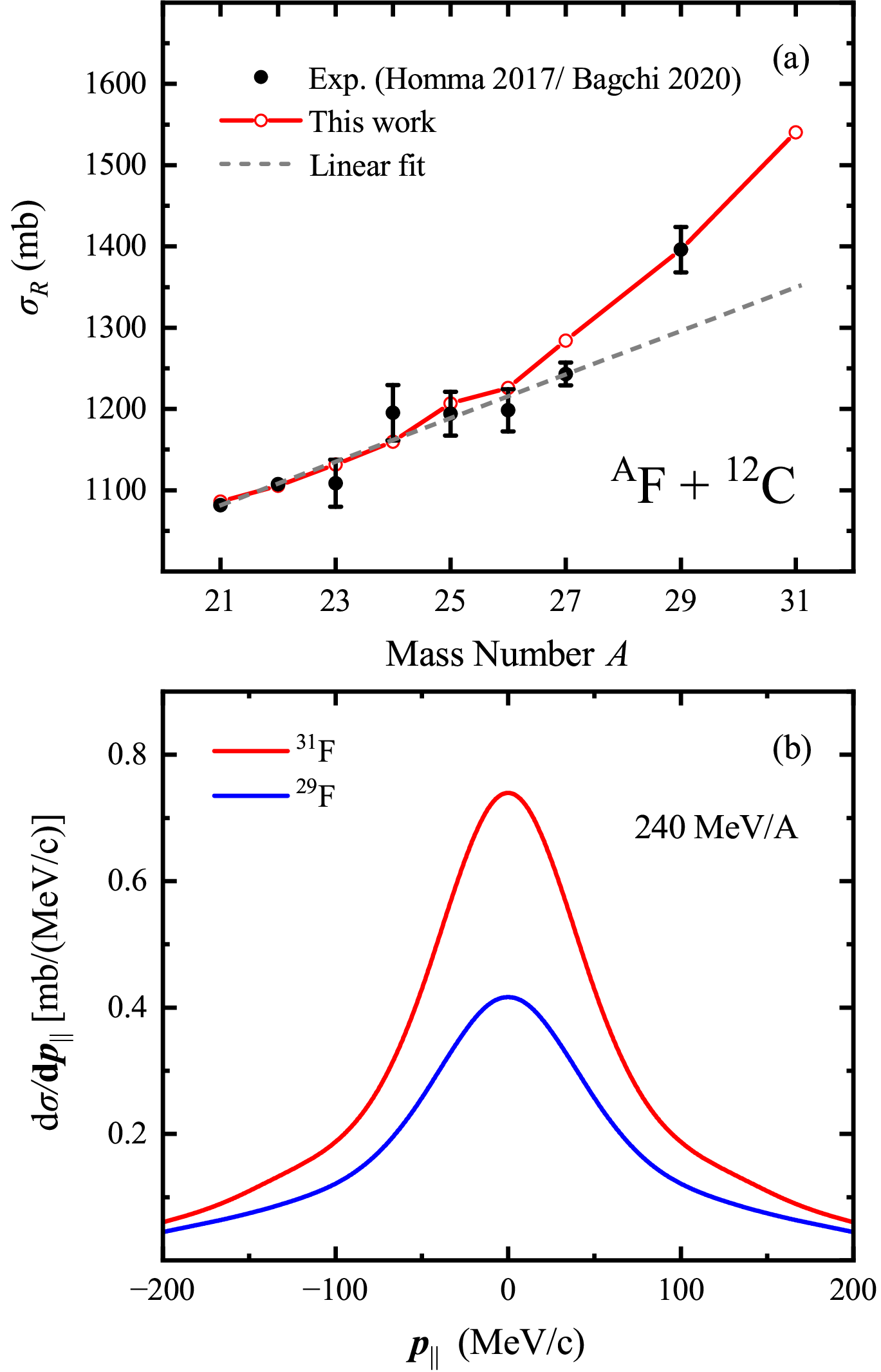}
  \caption{(Color online) (a) RCSs $\sigma_{R}$ of $^{21\text{--}31}$F + $^{12}$C at 240~MeV/A. 
  Open red points stand for our predictions with the Glauber model with inputs from the DRHBc theory. 
  For $^{29}$F + $^{12}$C, the RCS is obtained from a constrained DRHBc calculations with a commonly proposed deformation parameter $\beta_2 =$ 0.3. 
  Experimental data (closed black circles) are taken from Ref.~\cite{Bagchi2020PRL,homma2017measurements} for comparison.
 The dashed grey line shows the linear fit to the experimental data for $^{21\text{--}27}$F + $^{12}$C.
  (b) Inclusive longitudinal momentum distributions of residues after $2n$ removal reactions of $^{29,31}$F + $^{12}$C at 240~MeV/A. 
  }
  \label{fig:F}
\end{figure}

The calculated RCSs for fluorine isotopes on a carbon target at 240~MeV/A are shown in Fig.~\ref{fig:F} (a). 
Our results agree well with the available experimental data for $^{21\text{--}29}$F + $^{12}$C. The red line deviates a lot from the dashed grey line which is the linear fit of the experimental data except for $^{29}$F + $^{12}$C.
The slope of the RCSs from $^{27}$F to $^{29}$F and that from $^{29}$F to $^{31}$F are a least twice steeper than that from $^{25}$F to $^{27}$F on a carbon target, which also indicate possible halo structure in $^{29,31}$F.

In this scheme, the longitudinal momentum distributions for the residues after $2n$ removal reactions $^{29,31}$F + $^{12}$C at 240~MeV/A are displayed in Fig.~\ref{fig:F} (b), exhibiting pronounced narrow peaks for both reactions.
These are direct evidence for $2n$ halo structure in $^{29,31}$F.

\subsection{Application to $^{39}$Na on a carbon target}

A former DRHBc study~\cite{Zhang2023PRC(L1)} on the structure of $^{39}$Na has suggested that an oblate $2n$ surround a prolate core in a weakly bound halo nucleus with a small $S_{2n}$ ($\approx 1$~MeV). 
These structure information needs to be put into the Glauber model to predict the reaction observables for future experiments.
Similar as fluorine isotopes, we carry out the systematic study on the reaction observables of neutron-rich sodium isotopes on a carbon target. 

\begin{figure}[htbp]
  \centering
\includegraphics[width=0.8\linewidth]{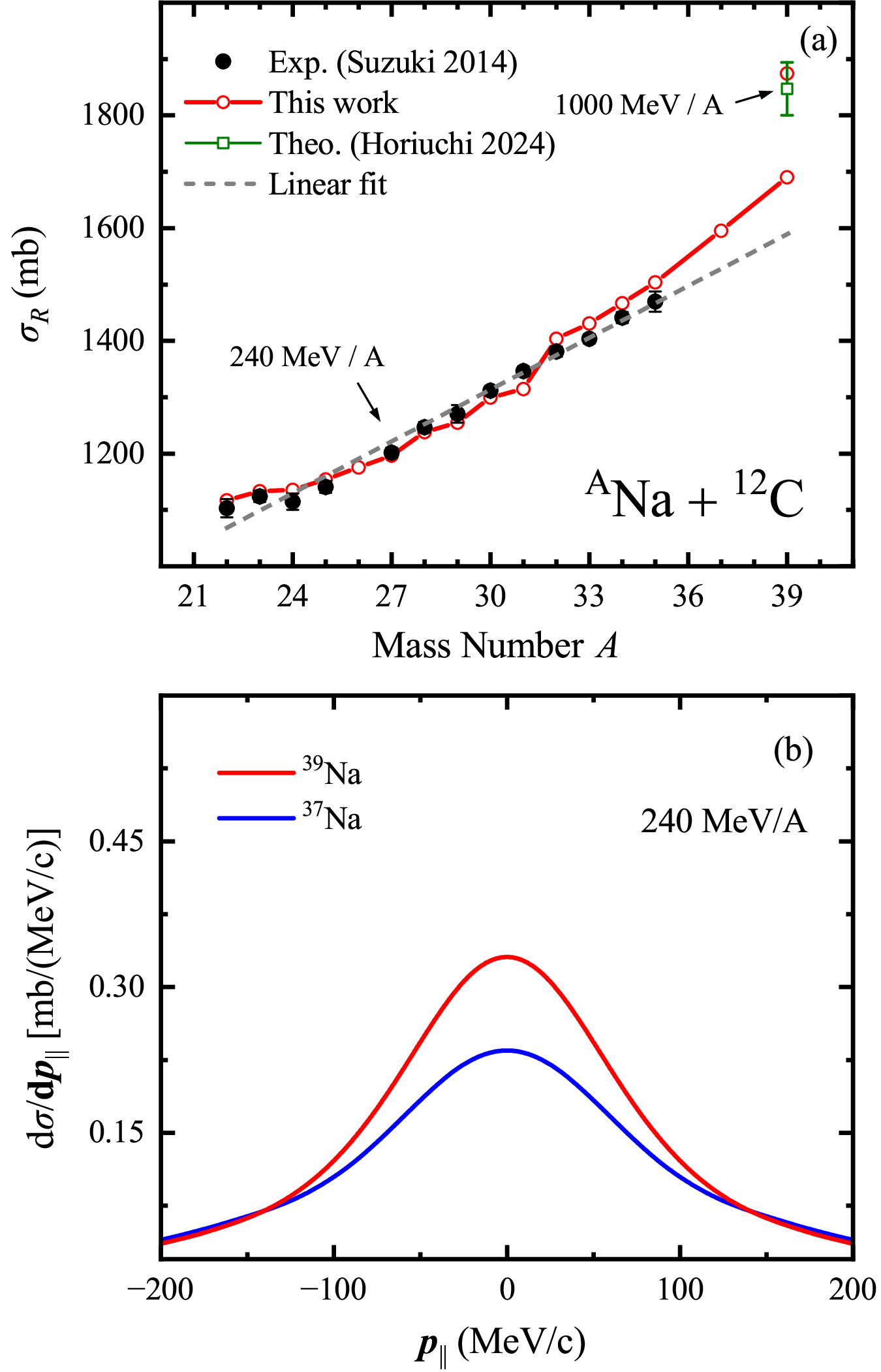}
  \caption{(Color online) 
  (a) RCSs $\sigma_{R}$ of $^{22\text{--}39}$Na + $^{12}$C at 240 and 1000~MeV/A. 
  Open red points stand for our predictions with the DRHBc + Glauber approach, in comparison with a linear fit (the dashed grey line) on available experimental data (closed black circles) for $^{22\text{--}35}$Na~\cite{Suzuki2014EPJ}.
  Open green square shows the result for $^{39}$Na + $^{12}$C at 1000~MeV/A calculated by Horiuchi~\cite{Horiuchi2024PLB}.
  (b) Same as Fig.~\ref{fig:F} (b), but for $^{37,39}$Na.}
  \label{fig:Na}
\end{figure}

The calculated RCSs for sodium isotopes on a carbon target are displayed in Fig.~\ref{fig:Na} (a). 
Our results at 240~MeV/A are in satisfactory agreement with the experimental data up to $^{35}$Na, which show the validity of the DRHBc + Glauber approach. 
A notable increase of the RCS (red open circle line) appear from $^{37}$Na to $^{39}$Na, compared with the  extrapolation of the linear fit on available data~\cite{Suzuki2014EPJ}. 
This deviation is due to a transition of the valence neutrons from the dominated $d$-orbital to spatially diffuse $p$-orbital, which reveals the essential mechanism for the formation of $p$-wave halo structure.
Furthermore, the consistency of our prediction at 1000~MeV/A with another independent theoretical study by Horiuchi \emph{et al.}~\cite{Horiuchi2024PLB} double confirm the reliability and capability of our model.

As a direct probe for the halo feature, we also calculate the longitudinal momentum distributions of the residues after $2n$ breakup reactions for $^{37,39}$Na + $^{12}$C. 
As plotted in Fig.~\ref{fig:Na} (b), the result for $^{39}$Na + $^{12}$C are narrower than that for $^{37}$Na + $^{12}$C, which is a pronounced feature for a halo nucleus. 
Quantitatively, the full width at half maximum (FWHM) for $^{37}$Na and 
$^{39}$Na are 182~MeV/c and 158~MeV/c, respectively.
Therefore, $^{39}$Na is a promising $2n$ halo nucleus from microscopic structure to reaction observables.

\section{Summary}\label{sec:Summary} 
In all, we check the validity of the Glauber model for $2n$ halo nucleus $^{11}$Li first. 
Then, combining the Glauber model with the DRHBc theory, we complete a thorough study on neutron-rich fluorine and sodium isotopes from microscopic structure to reaction observables.

Our calculations of the RCSs for the fluorine and sodium isotopes agree well with the available experimental data. 
Sudden increase of the RCSs and narrow longitudinal momentum distributions support $^{29,31}$F and $^{39}$Na as the promising candidates with dilute $2n$ halo structure.

As the benchmark work of $2n$ halo description, the DRHBc + Glauber approach has been build up for the first time, and successfully applied to provide quantitative quantities of $2n$ halo candidates $^{31}$F and $^{39}$Na for future measurements. This casts a new light on the search for new halo nuclei in the medium-mass region of $A\approx 40$.

\section*{Acknowledgment}
\indent This work was supported by the National Natural Science Foundation of China (Grant Nos.~12575122, 12175010, 12305125) and the National Key Laboratory of Neutron Science and Technology (Grant No. NST202401016).

\bibliography{F_Na}

\end{document}